\documentclass{aa}
\usepackage{txfonts,natbib}
\usepackage{float,epsfig}
\usepackage{pstricks}

\newcommand{\mincir}{\raise-2.truept\hbox{\rlap{\hbox{$\sim$}}\raise5.truept \hbox{$<$}\ }}
\newcommand{\magcir}{\raise-2.truept\hbox{\rlap{\hbox{$\sim$}}\raise5.truept \hbox{$>$}\ }}

\begin{document}

\title{Hydrodynamic simulations of correlation and scatter in galaxy cluster maps.}

\author{A. Finoguenov\inst{1,2} \and A.~J.~R. Sanderson\inst{3} \and J.~J. Mohr\inst{4,1}, J.J. Bialek\inst{6} \and A. Evrard\inst{6}}

\institute{
Max-Planck-Institut f\"ur extraterrestrische Physik,
             Giessenbachstra\ss e, D-85740 Garching, Germany \and
University of Maryland Baltimore County, 1000 Hilltop circle, Baltimore, MD
21250, USA \and 
School of Physics and Astronomy, University of Birmingham, Edgbaston,
Birmingham, B15 2TT, UK \and
Department of Physics,
     Ludwig-Maximilians-Universitaet, Scheiner Strasse 1, D-81679
     Munich, Germany
\and
Physics Department, University of Michigan, Ann Arbor, MI 48109, USA; Astronomy Department, University of Michigan, Ann Arbor, MI 48109, USA; Michigan Centre for Theoretical Physics, Ann Arbor, MI 48109, USA
}

\date{Received Jan. 12 2009; accepted Nov. 24 2009}

\abstract{} {The two dimensional structure of hot gas in galaxy clusters
  contains information about the hydrodynamical state of the cluster, which
  can be used to understand the origin of scatter in the thermodynamical
  properties of the gas, and to improve the use of clusters to probe
  cosmology.}  {Using a set of hydrodynamical simulations, we provide a
  comparison between various maps currently employed in the X-ray analysis
  of merging clusters and those cluster maps anticipated from forthcoming
  observations of the thermal Sunyaev-Zel'dovich effect.}{We show the
  following: 1) an X-ray pseudo-pressure, defined as square root of the soft
  band X-ray image times the temperature map is a good proxy for the SZ map;
  2) we find that clumpiness is the main reason for deviation between X-ray
  pseudo-pressure and SZ maps; 3) the level of clumpiness can be well
  characterized by X-ray pseudo-entropy maps. 4) We describe the frequency
  of deviation in various maps of clusters as a function of the amplitude of
  the deviation. This enables both a comparison to observations and a
  comparison to effects of introduction of complex physical processes into
  simulation.}{}


\maketitle

\section{Introduction}

The X-ray emission from clusters of galaxies has been conventionally
characterized in terms of an X-ray surface brightness image and a
temperature map.  Historically, due to the success of the Einstein (Gursky
et al. 1972) and ROSAT (Boehringer et al. 2000, 2001) missions, X-ray
imaging data are available for a wide range of clusters, both in mass and
redshift, while cluster searches based on the level of X-ray emission have
become one of the primary cosmological tools (for a review see Borgani
\& Guzzo 2001). 

The availability of cluster temperatures, on the other hand, has long been
limited to single emission-weighted values, as provided by Ginga, Einstein
and EXOSAT (David et al. 1993; Edge et al. 1992), and although the first
temperature maps were made with ROSAT (Briel \& Henry 1994) and ASCA
(Markevitch 1996, 1999) data, they were limited in either spectral range
(ROSAT) or spatial resolution (ASCA).  With the advent of Chandra and
XMM-Newton, temperature maps have become widely available, revealing an
unprecedented amount of detail (Vikhlinin et al.  2001; Mazzotta et al 2002;
Briel et al. 2004; Finoguenov et al. 2004; Henry et al. 2004).  However,
neither temperature maps nor X-ray images are the primary characteristics of
the gas, and are instead results of underlying pressure and entropy
distribution, that trace the dark matter potential, reflects the
thermodynamical history of the gas and the on-going magneto-hydrodynamical
processes. However, projection effects complicate this study and in practice
X-ray pseudo pressure and entropy maps have been introduced (Markevitch et
al.  1996; Finoguenov et al.  2004, 2005, 2006a,b, 2007; Briel et al. 2004;
Henry et al. 2004), addressing an immediate goal of discriminating between
the various origins of fluctuations, e.g. to differentiate between shocks
and gas displacement, also known as cold fronts (Vikhlinin et al. 2001).
The way these maps work is that at the position of the cold front the
contrast in the entropy map is increased, while the contrast in the pressure
map is removed, compared to either images or temperature maps, while on the
positions of the weak shocks, the contrast in the entropy maps is
suppressed, while the contrast on the pressure map is enhanced. Some
astonishing discoveries using these maps were made using very deep XMM
observations of M87, recovering the details of AGN feedback (Simionescu et
al. 2007) and the underlying triaxiality of the dark matter distribution.

\begin{figure*}
\includegraphics[height=18cm]{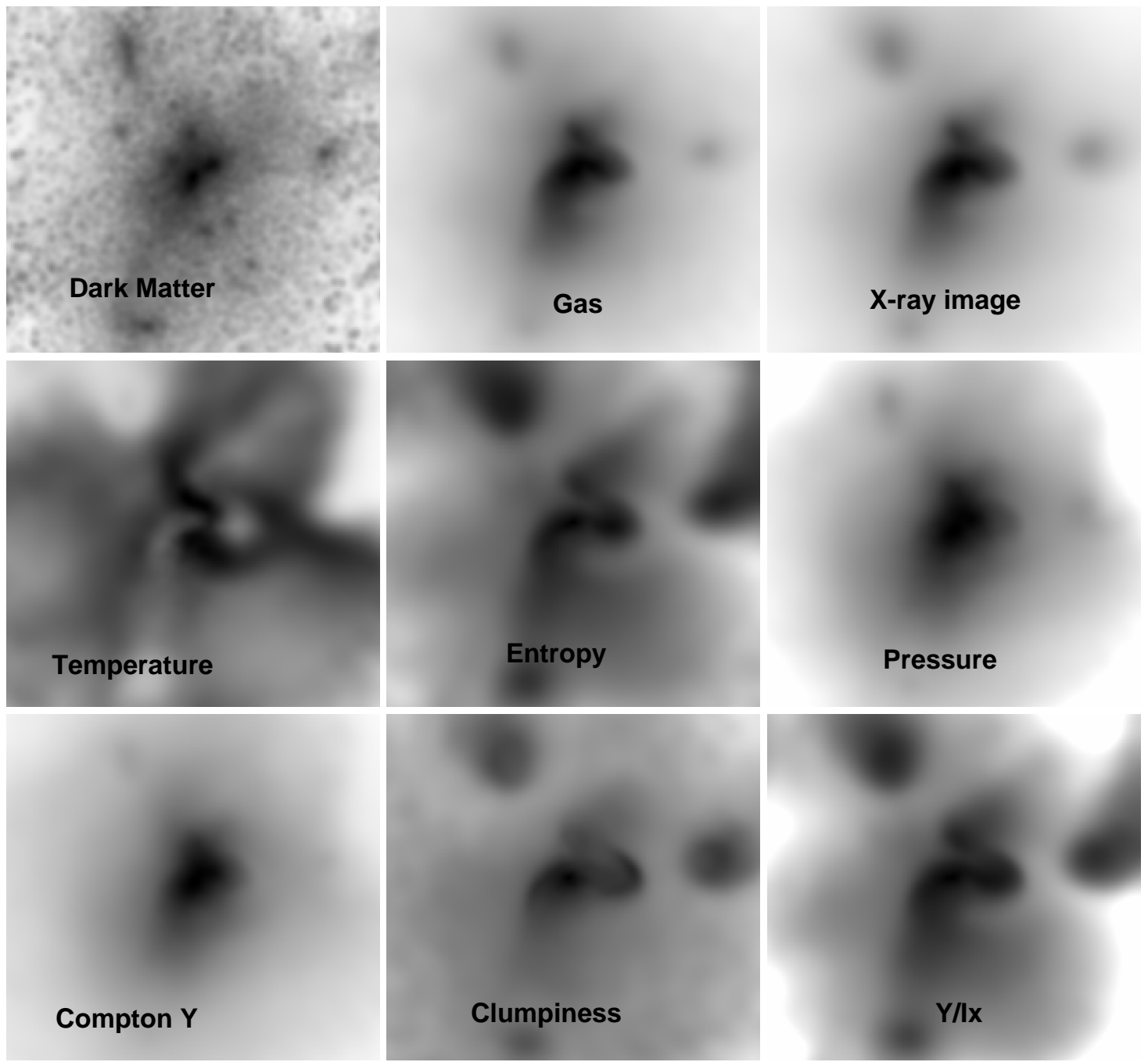}
\caption{Example of the cluster maps in one of the simulation runs. From
  the top left to the bottom right we show the dark matter column density,
  gas column density, X-ray images in the 0.5--2 keV band, X-ray temperature
  map, pseudo entropy map, pseudo pressure map, Compton Y map, clumpiness
  map, ratio of Compton Y to X-ray image. The grey scale is set to depict
  the central structure in black in all the maps.
\label{f:ex}}
\end{figure*}

These pressure and entropy distributions are called pseudo maps to reflect
the fact that both pressure and entropy are local quantities, while the
projected emission weighted temperature and X-ray surface brightness are
averaged along the line of sight. The effect of projection in reducing the
fluctuation has been found to be a function of the spatial scale of
fluctuation (Schuecker et al. 2004), with small-scale fluctuations being
more strongly suppressed. Since all cluster maps suffer from projection
effects, it is important to understand the link between various methods of
studying clusters, such as thermal SZ, weak lensing and X-ray.

In order to improve our understanding of cluster physics, we employ a set of
hydrodynamical simulations to study how well various cluster maps compare.
We paid a particular attention on possible use of cluster maps for
diagnostics not readily available to the observer. In \S2, we describe the
simulation sample. In \S3 we look for correlation among the maps, discuss
the general trends and look for correlations of deviations from the general
trends. We also present the statistical expectation for deviations to occur
as a function of their strength. We summarize our findings in \S4.

\section{Mock Observations}\label{s:mock_obs}

We use a catalog of 68 galaxy clusters evolved with the P3MSPH code (Evrard
1988, 1990), incorporating preheating of the ICM to mimic the effects of
galaxy feedback.  The level of entropy introduced into the initial
conditions is tuned to $106$~keV~cm$^2$ in order to match the observed
scalings of luminosity and ICM mass with temperature (Bialek et al. 2001).
No additional AGN feedback and radiative cooling has been included into the
simulations.

\begin{figure*}
\mbox{\includegraphics[width=8cm]{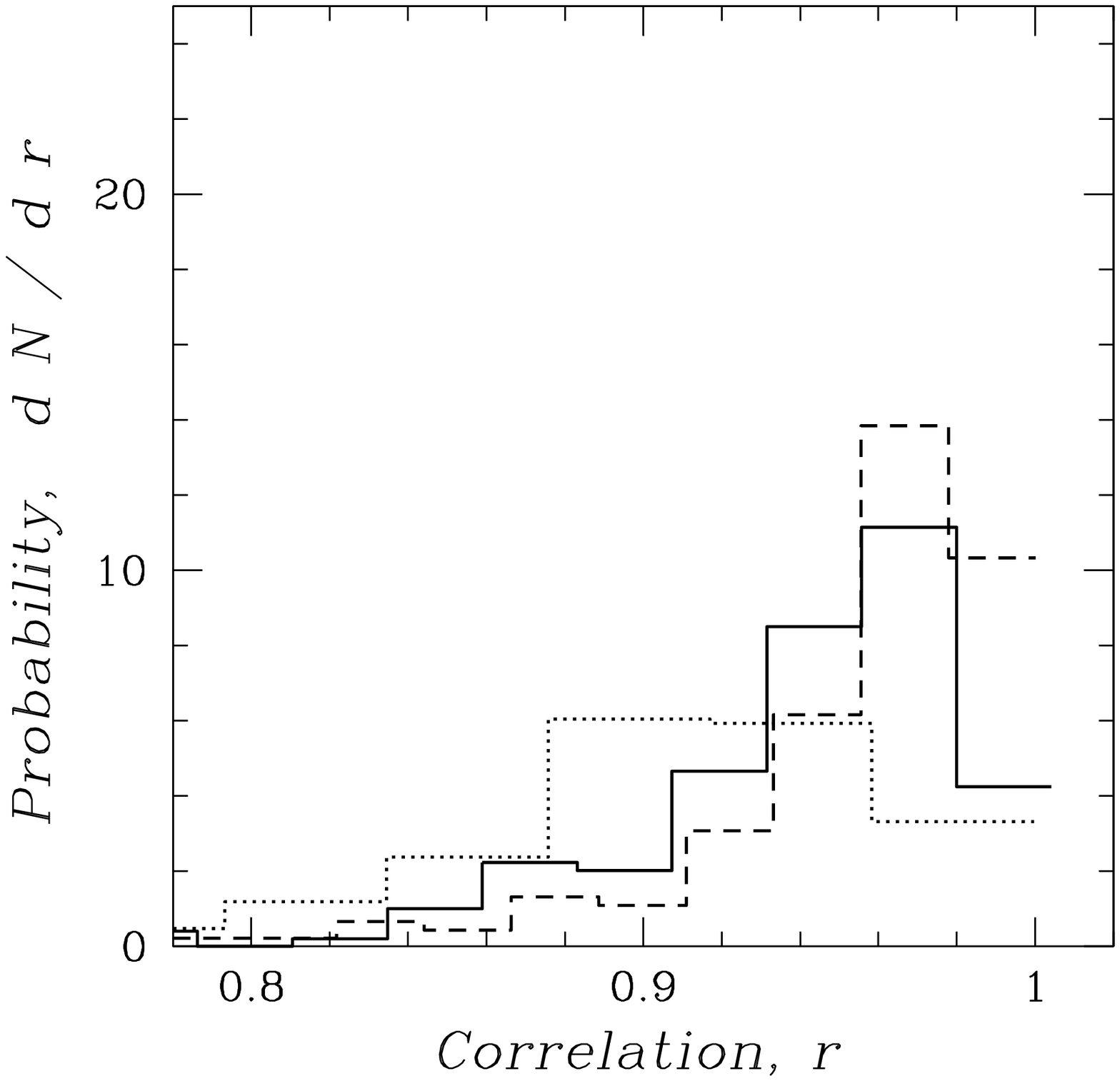}\hfill
\includegraphics[width=8cm]{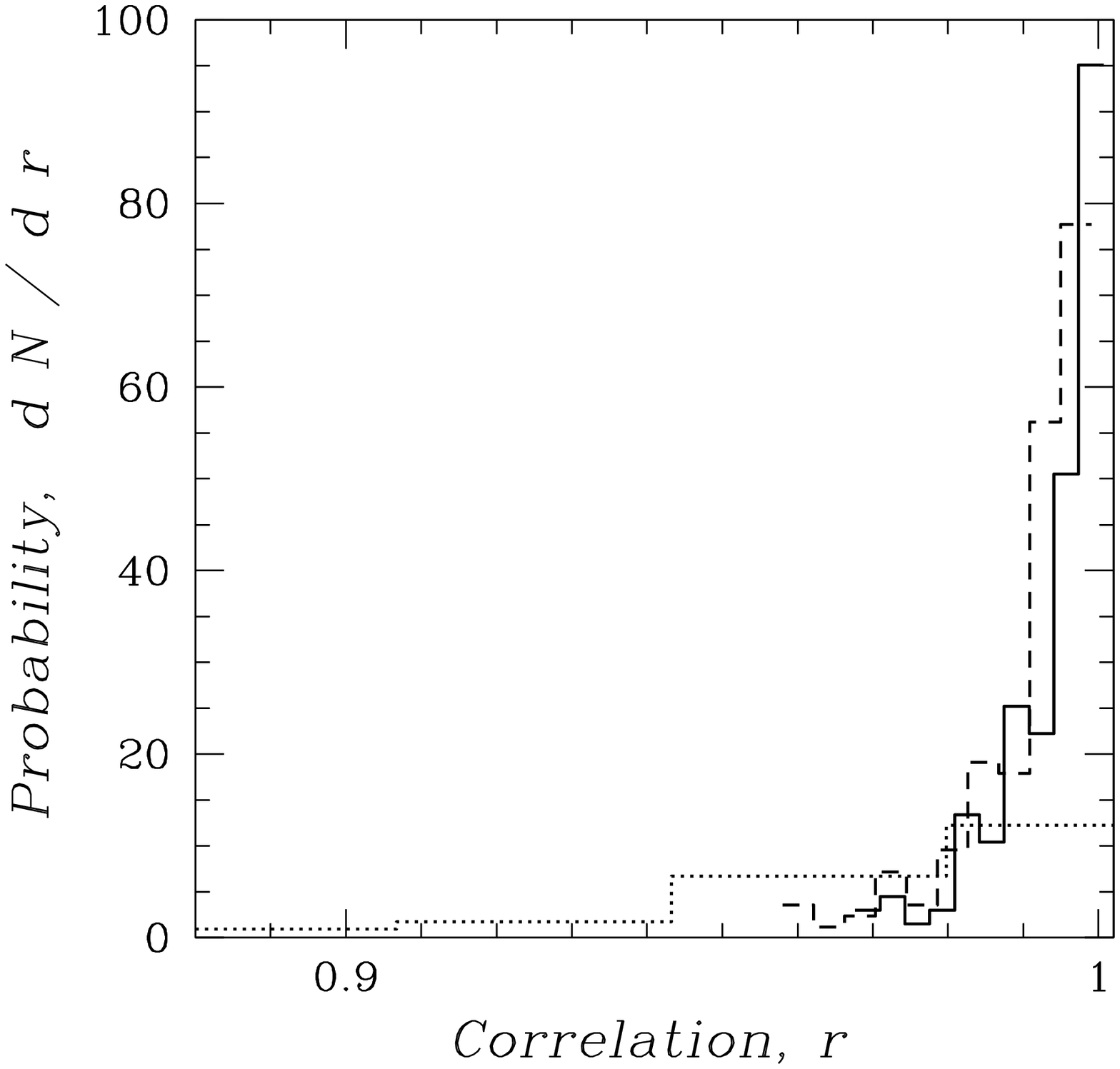}} 
\caption{The distribution of correlation coefficients between various 
simulated maps for each of the 68 simulated clusters, calculated using
Equation~\ref{eqn:corr}. {\it Left} panel represents correlations between
dark matter column density and each of the pressure (solid), Compton Y
(dashed) and square root of the X-ray surface brightness $I_X^{0.5}$
(dotted). {\it Right} panel represents correlations between Compton Y and
each of the pressure (solid) and $I_X^{0.5}$ (dotted).  The dashed line
shows a correlation between the ratio of Compton Y and X-ray surface
brightness with pseudo entropy. All comparisons are carried out
within 1/4 of $r_{200}$.
\label{f:sr}}
\end{figure*}

The simulations were produced using a multi-step procedure outlined in
(Bialek et al. 2001).  A single cluster from this dataset was found to
exhibit a post-merger cold front/features driven by sub-halo sloshing, as
presented in Bialek, Evrard \& Mohr (2002) and further detailed in Pawl et
al.  (2005).  The simulations evolve isolated cut-outs of a larger N-body
realization, with regions chosen such that final halo masses are sampled
with roughly equal likelihood across the mass range $M_{200} = 0.2-2 \times
10^{15} M_\odot$.  The runs are of moderate resolution, with gravitational
softenings between 50 and 150 kpc, and dark matter particle masses of $0.4-4
\times 10^{10} M_\odot$. While Lewis etal (2000) caution that large
softening can affect the overall cluster potential, the comparison study of
Frenk etal (1999) demonstrates that moderate resolution runs offer good
convergence in dark matter and gas properties on scales larger than roughly
twice the hydrodynamic smoothing scale.

A typical halo is resolved by $\sim 50,000$ particles.  With preheating of
$106$ keV cm$^{-2}$, this is sufficient to resolve low-order measures of ICM
properties, such as temperature and luminosity, and morphological features
driven by mergers with mass ratios of order $0.1$ and larger.  The
underlying cosmology is a flat, concordance model with $\Omega_{m} = 0.3$,
$\Omega_{\Lambda} = 0.7$, $\Omega_{b} = 0.03$, $\sigma_8 = 1.0$, and $h =
0.7$, where the Hubble constant is defined as $100h$~km~s$^{-1}$~Mpc$^{-1}$
and $\sigma_8$ is the power spectrum normalization on $8h^{-1}$ Mpc scales.
Details of the full ensemble are presented on the project
web-page\footnotemark[1].

In this paper we only used the final (z=0) configuration of 68 model
clusters and employ 3 orthogonal projections of each model as independent
data, giving a total of 204 separate datasets.  Every map has 256 pixels on
a side, spanning a region twice the virial radius of the system. By
construction the maps are designed to study the appearance of individual
clusters and not the projected effects in cluster surveys (e.g. White 2003).
From an initial set of X-ray surface brightness (I) and temperature maps
(T), we have constructed the X-ray pseudo pressure and entropy maps
(hereafter P and S, respectively) as $P=I^{1/2}*T$ and $S=T/I^{1/3}$ and
compared those to the column density of the both ionized gas (G) and dark
matter (DM), as well as the predicted Compton $Y$ parameter of the thermal
Sunyaev-Zel'dovich (SZ; Sunyaev \& Zel'dovich 1980) effect (Y). We define
the clumpiness as $C = G / \sqrt{I}$, which differs from its exact
definition by the square root of the projection length. In Fig.~\ref{f:ex}
we show the simulated cluster maps obtained for the g1 run, 1st projection.

The dark matter maps in simulations appear much clumpier compared to the
gas. This is likely due to erosion of small-scale structure in the gas by
setting the entropy floor. A careful comparison between dark matter subhalos
and the hot gas has been made in Powell et al. (2009).  Using the case of
one halo simulated with and without cooling, they show that occupation of
small subhalos by hot gas is efficiently suppressed under both treatments.
Dark matter maps of cluster are thus likely to appear clumpier than the
X-ray gas irrespective of the detailed gas physics assumptions.  Near
equilibrium, the gas will trace the equipotential surfaces, which tend to be
smoother and rounder than the density isosurfaces. In order to remove the
small-scale power in the comparison, we have convolved the dark matter maps
with the Gaussian.  The members of the simulated sample range in spectral
temperature from 1.5 keV to about 8 keV, with cluster masses $M_{200}$
ranging from (0.015--2.4)\,$\times 10^{15} M_{\odot}$. The derived images,
along with associated parameters describing the simulations are publicly
available as part of VCE, the {\it Virtual Cluster
  Exploratory}\footnotemark[1].
\footnotetext[1]{http://vce.physics.lsa.umich.edu}

\section{Results}

For each simulated cluster we can quantify the correlation between the
different maps within a fixed fraction of the virial radius. We use the
following expression for the correlation coefficient between images $X$ and
$Y$ for a given cluster

\begin{equation} \label{eqn:corr}
r = {\sum_i ( X_i-\overline{X}) ( Y_i- \overline{Y}) \over \sqrt{\sum_i ( X_i- \overline{X})^2 ( Y_i- \overline{Y})^2 }} ,
\end{equation}

where $i$ sums over all the pixels in the image. We search for correlations
between all primary maps. The distribution of correlation coefficients
across the ensemble of simulated clusters is shown in Fig.~\ref{f:sr}. The
strongest correlation occurs between P and Y. The dark matter is equally
well correlated with either Y or P, and somewhat less well with I.

We have found strong correlations between X-ray pseudo-entropy and the ratio
of Y to I; which we attribute to the fact that the entropy can also be
defined as $Y / I^{5/6}$. This can useful in case where only X-ray imaging
is available as well as good SZ-data, but no temperature map, as should be
the case for high-z clusters. Historically, the ratio of Y-to-I was proposed
for measurement of the Hubble constant (see e.g. Bonamente et al. 2007). Our
results provide an insight on the origin of scatter in this measurement, and
our proposal to reduce the scatter in this comparison consists of the use of
pseudo pressure maps instead of X-ray images and also estimating the
expected deviation based on the amplitude of the scatter seen in the pseudo
entropy maps.

\subsection{Estimated influence of the general trend}

In this section we estimate analytically the effect of projection on the
derived correlation coefficient, accounting therefore for differences in the
radial trends between the maps and compare these predictions to the observed
correlation coefficients.

Consider the scaling with radius of different properties.

\begin{eqnarray}
n \propto (1+(r/r_c)^2)^{-3 \beta /2 }\\
\beta=2/3\\
S  \propto (1+(r/r_c)^2)^{0.55}\\
\longrightarrow T \propto n^{0.12}\\
\longrightarrow I \propto (1+(r/r_c)^2)^{-3\beta + 0.5 } \propto
(1+(r/r_c)^2)^{-1.5}\\
\longrightarrow I^{0.5} \propto (1+(r/r_c)^2)^{-0.75}\\
\longrightarrow P =  I^{0.5}T \propto (1+(r/r_c)^2)^{-0.87}\\
\longrightarrow Y \propto (1+(r/r_c)^2)^{-3\beta /2 \times 1.12 + 0.5 } \propto (1+(r/r_c)^2)^{-0.62}\\
DM \propto (1+(r/r_c)^2)^{-1.3+0.5}  \propto
(1+(r/r_c)^2)^{-0.8}
\end{eqnarray}

where n is the gas density, $\beta$ is the radial slope for which we assumed
a typical cluster value. The entropy profile is assumed, based on the
results of observations (e.g. Finoguenov et al. 2005). The DM is the
projected column density of dark matter, for which the distribution was
taken to resemble closely the NFW profile (Navarro, Frenk \& White 1997).
Other parameters can be derived based on these assumptions. Addition of the
constant 0.5 is a result of the Abell integral.

\begin{figure}[ht]
\includegraphics[width=8cm]{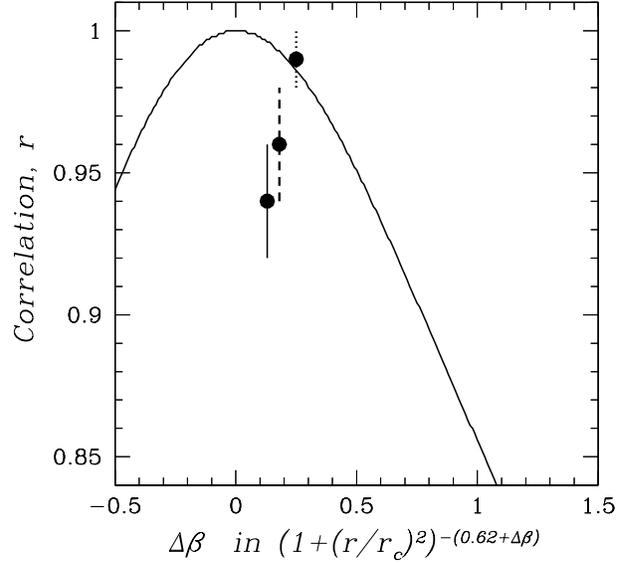}
\caption{Variation in the correlation coefficient as a function of change
in the radial slope. The changes are calculated in respect to the Compton Y
two-dimensional distribution with the profile of
$(1+(r/r_c)^2)^{-(0.62+\Delta \beta )}$. The points show the actually
measured correlation strength and the range enclosing 68\% of the data. The
dotted, dashed and solid lines represent a correlation between Compton Y and
each of the pseudo pressure, the dark matter column density and the square
root of X-ray surface brightness ($I_X^{0.5}$), respectively. 
While the pseudo pressure and Compton Y parameter exhibit a correlation
consistent with the predicted trend, the correlation with both dark matter
and the X-ray image is lower than the
expectation based merely on differences in the profiles.
\label{f:beta}}
\end{figure}

In order to illustrate the effect of differences in the radial slope on the
observed correlation, in Fig..~\ref{f:beta} we calculated an expected
reduction in the correlation coefficient, assuming
$(1+(r/r_c)^2)^{-(0.62+\Delta \beta )}$, varying $\Delta \beta$ to cover the
range of all the observed clusters. The main conclusion of this test is that
while the typical correlation coefficients between Compton Y vs.  pseudo
pressure are reproduced, the observed correlations between simulated X-ray
image and Compton Y as well as all the correlation of Compton Y with dark
matter column density are much weaker. This implies both that the gas is
only an approximate tracer of the dark matter and that hydrodynamical
processes involving gas displacements (cold fronts) are playing a dominant
role in defining the soft X-ray band images of the cluster, in accordance
with recent Chandra and XMM results (Vikhlinin et al. 2001, Mazzotta et al.
2002; Briel et al. 2004) and other numerical simulations (Powell et al.
2009).

\subsection{Definition of substructure as deviation from the mean trend}

In order to see the correlation in the substructure on the images, we have
corrected for the radial variation in all maps. We perform this correction
by calculating an average map for each quantity, across the ensemble of
simulated clusters, which captures the broad systematic trend.  This average
map is then used to normalize each individual cluster map, to reveal the
deviations about the global trend caused by interesting cluster physics. The
correlation between these corrected maps then more cleanly probes the
relation between these deviations.

\begin{figure*}
\includegraphics[width=6.cm]{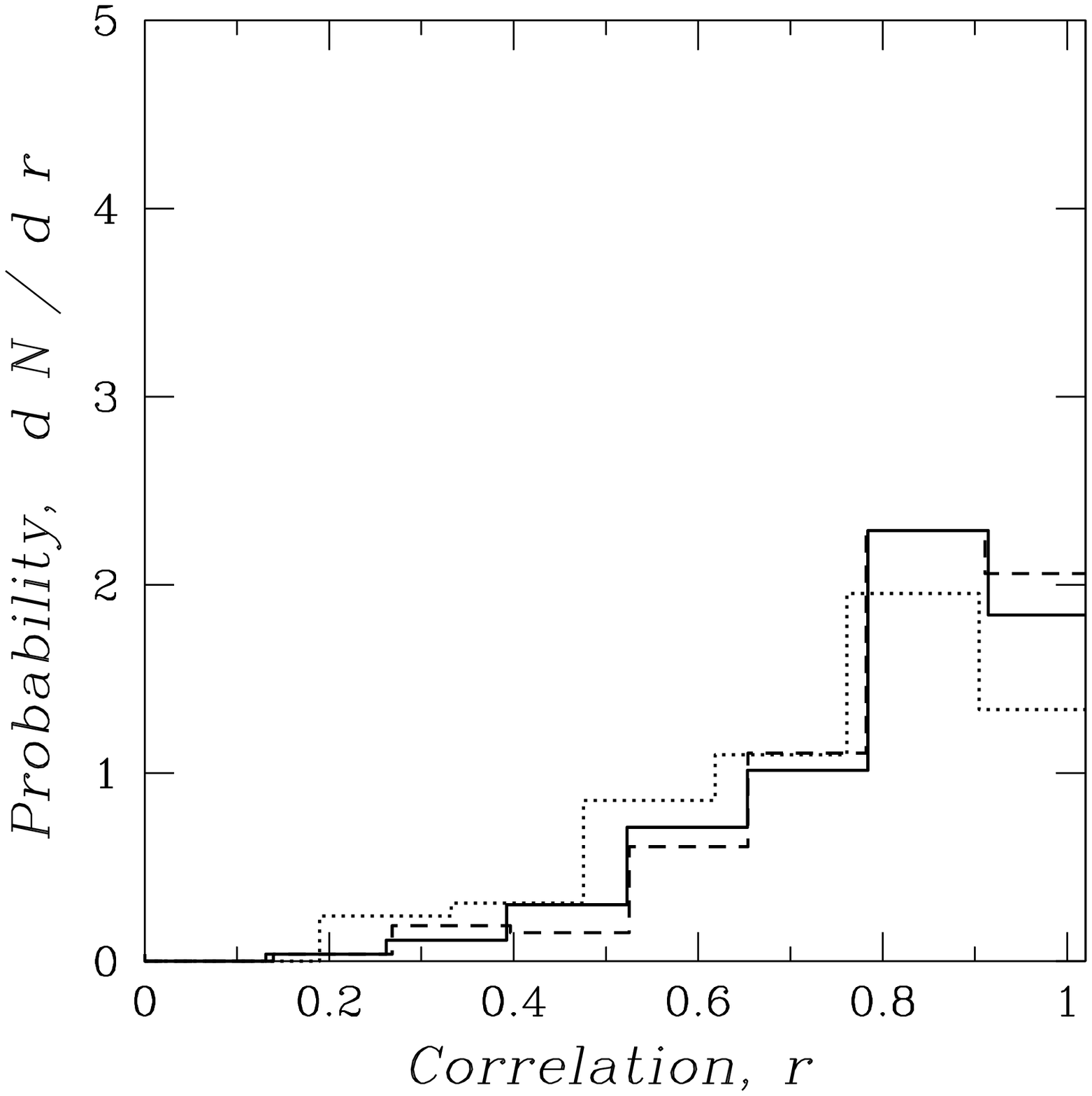}\hfill
\includegraphics[width=6.cm]{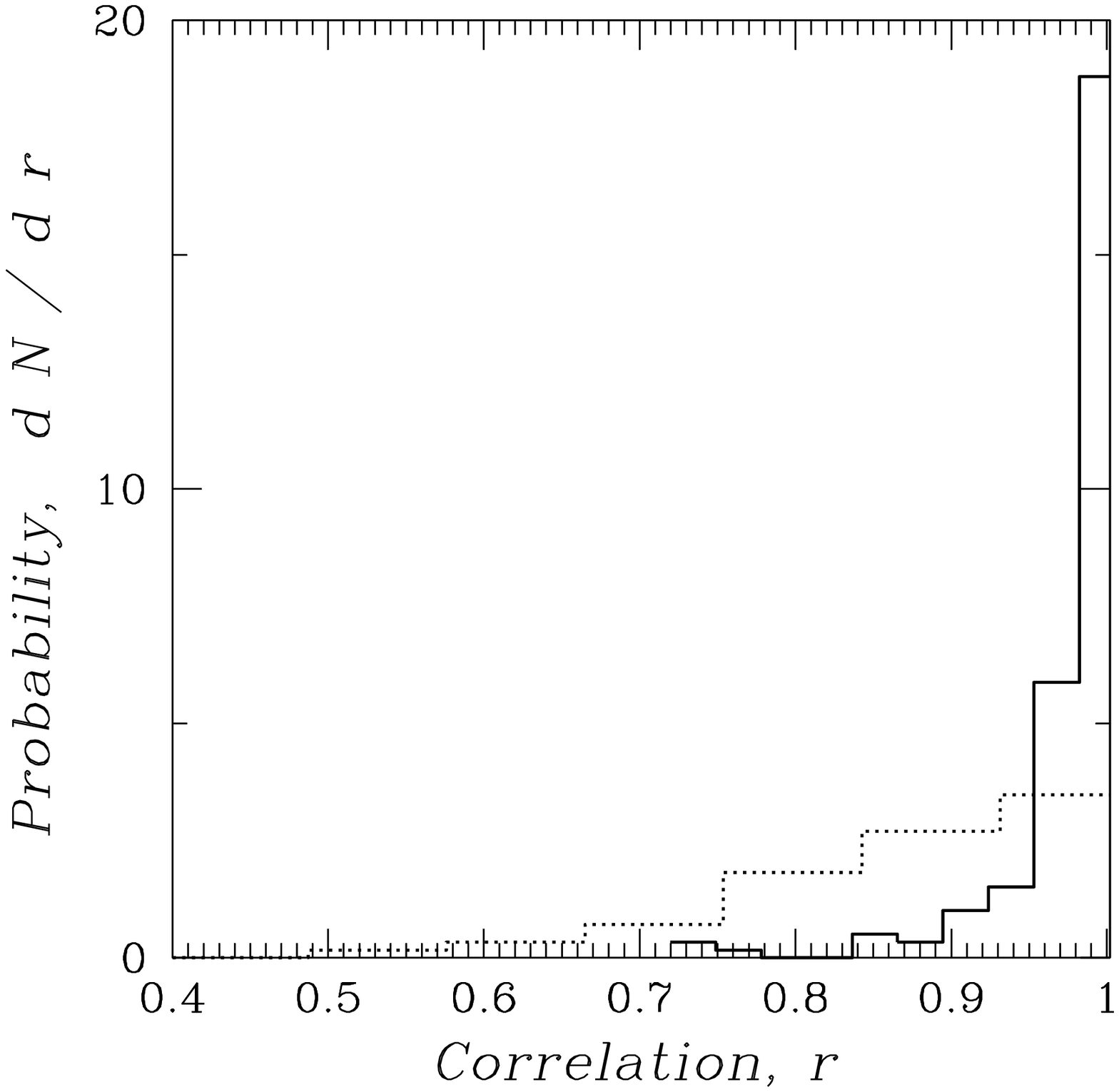}\hfill
\includegraphics[width=6.cm]{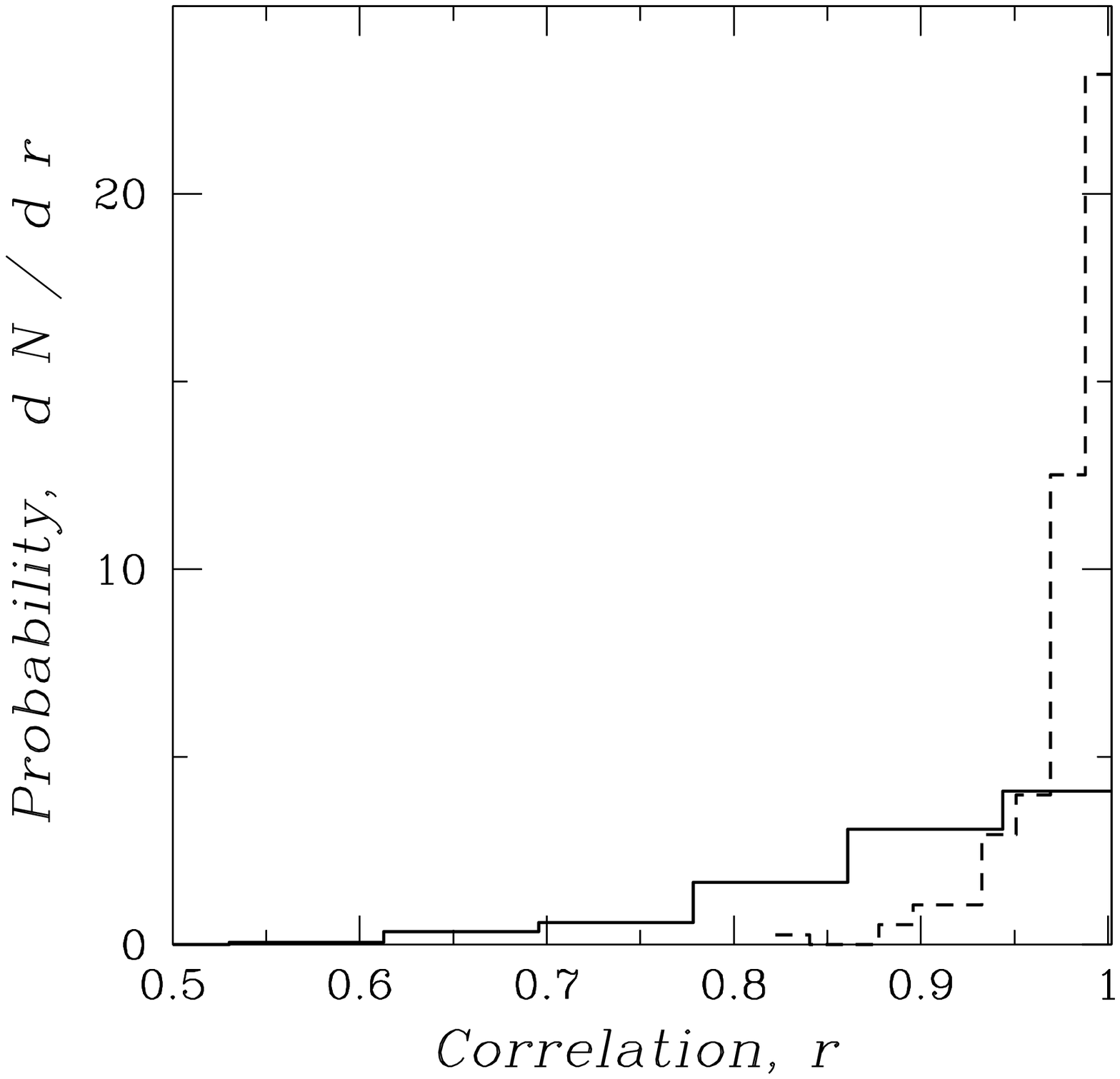}
\caption{The distribution of the pixel-by-pixel correlation coefficient
  (Equation~\ref{eqn:corr}) between the maps, after correcting for the mean
  trend in each map. {\it Left} panel represents correlations between dark
  matter column density and each of the pressure (solid), Compton Y (dashed)
  and square root of the X-ray surface brightness $I_X^{0.5}$ (dotted). {\it
    Middle} panel represents correlations between Compton Y and pressure
  (solid) and $I_X^{0.5}$ (dotted).  {\it Right} panel represents
  correlations between the entropy and the clumpiness (solid line) and
  correlations between the ratio of Compton Y and X-ray surface brightness with
  entropy (dashed).  All comparisons are carried out within 1/2 of
  $r_{200}$. Changing the radii of the comparison to 1/4 $r_{200}$ does not
  change these results.
  \label{f:csr}}
\end{figure*}

To eliminate bias from anomalous clusters, we identify and exclude outliers
from the calculation of the average map using the following iterative
scheme. Firstly, we take the logarithm of all the pixel values in the map
for every cluster and calculate the mean of these logged images. We then
compute the correlation coefficient between this mean logged image and the
individual logged map for each cluster. Any clusters with a correlation
below 0.85 are identified as outliers and the mean logged image is
recalculated with them excluded; the process is repeated until no further
clusters are excluded. This process is found to change the resultant map by
only $\sim$5\%. The distribution of correlation coefficients for the
corrected maps is plotted in Fig.~\ref{f:csr} and a comparison with
Fig.~\ref{f:sr} shows the much broader spread of values obtained when the
mean trend is removed. Intuitively, the level of correlation between the
gas-related maps and dark matter maps should be quite sensitive to the
physics introduced in simulations, such as cooling, conduction and
viscosity. Thus a comparison of predictions of various simulations to the
observational data shall be rewarding. The correlation between the
gas-tracing maps has also decreased. For example a correlation between the
pseudo-pressure to Compton Y maps dropped from 0.98 to 0.95. In order to
understand the origin of this decline we move to our next test - a direct
comparison of the levels of fluctuations of each cluster.

\subsection{Scatter}

\begin{figure*}
\mbox{
\includegraphics[width=4.5cm]{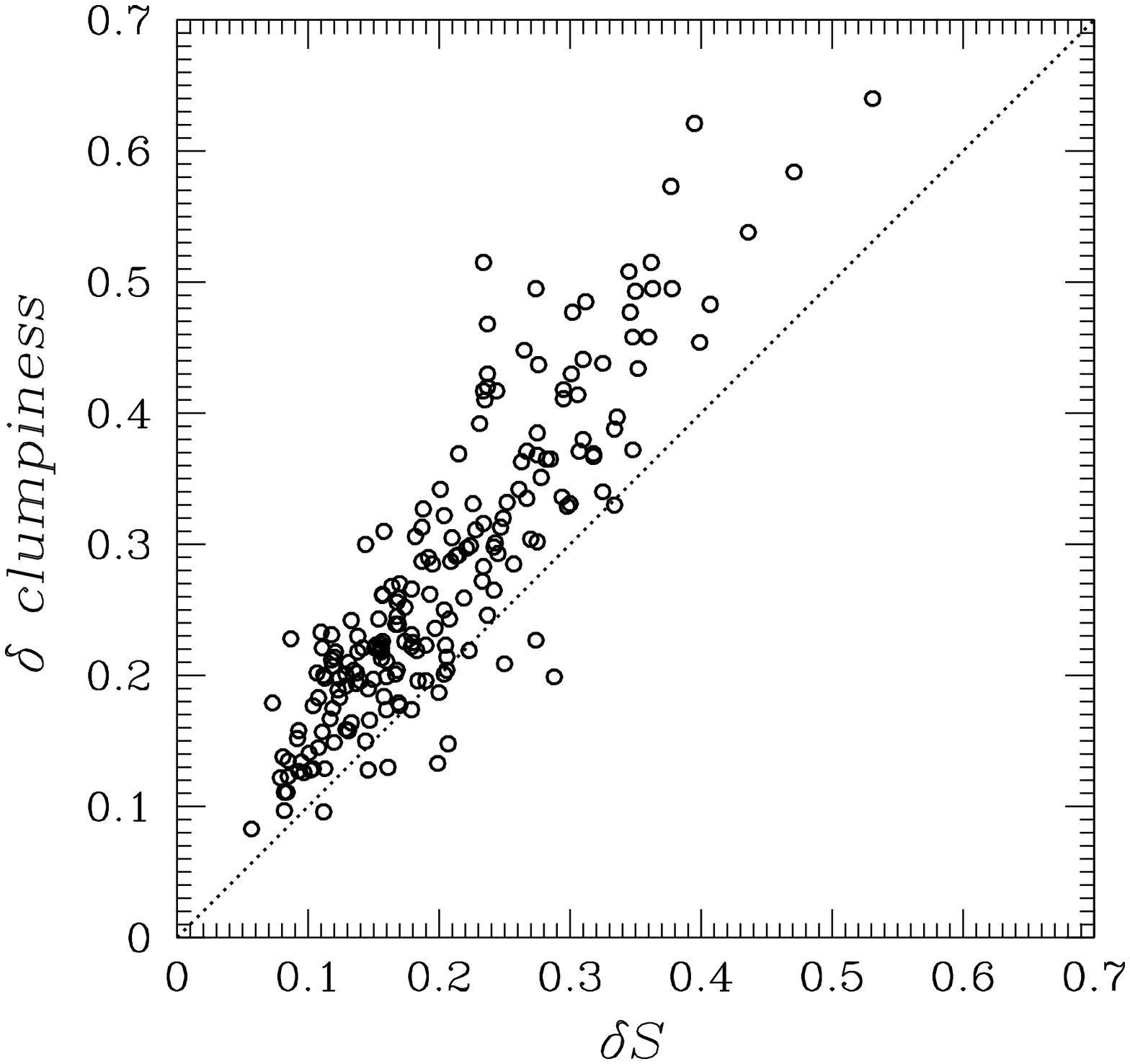}
\includegraphics[width=4.5cm]{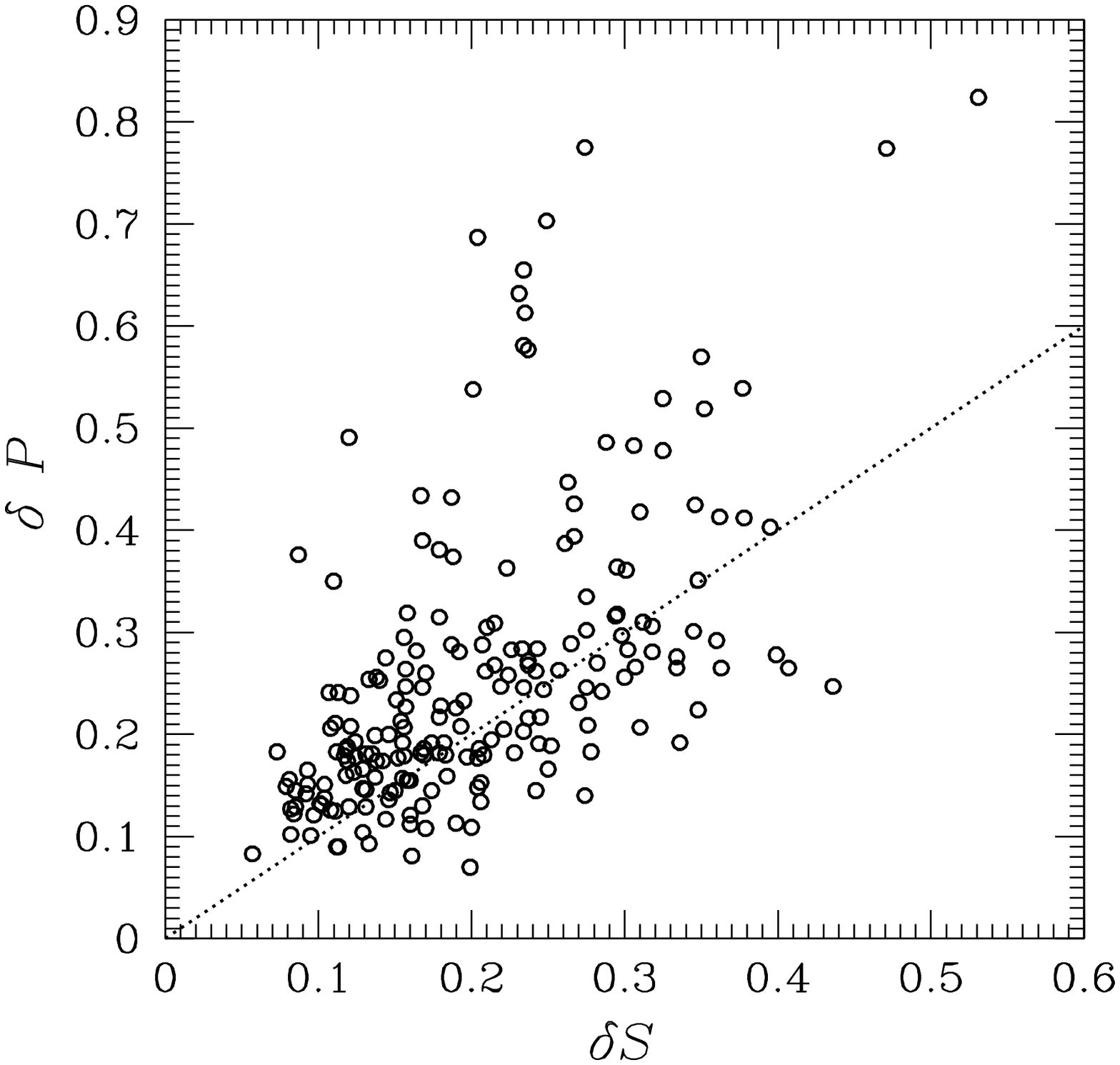}\hfill
\includegraphics[width=4.5cm]{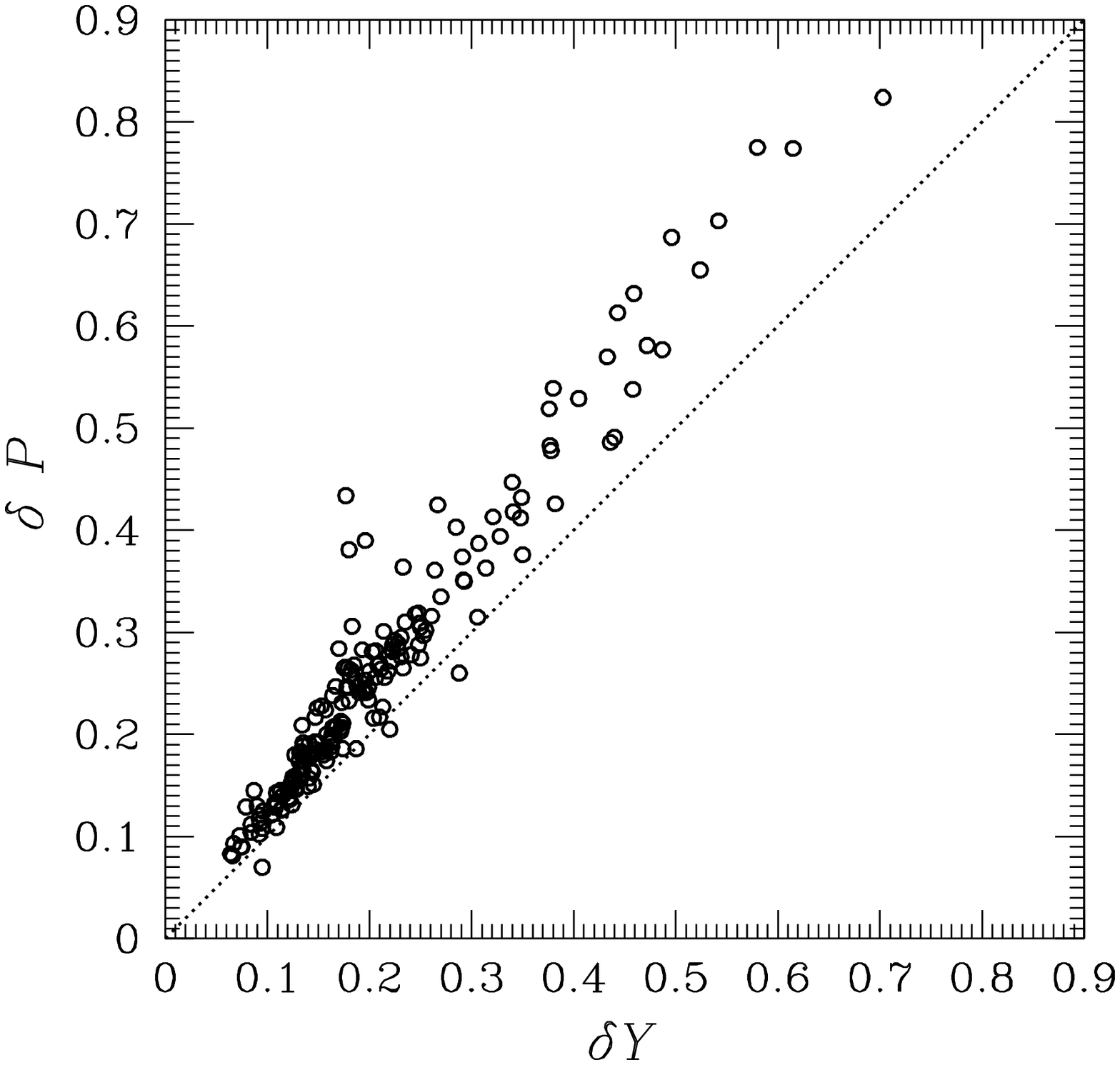}\hfill
\includegraphics[width=4.5cm]{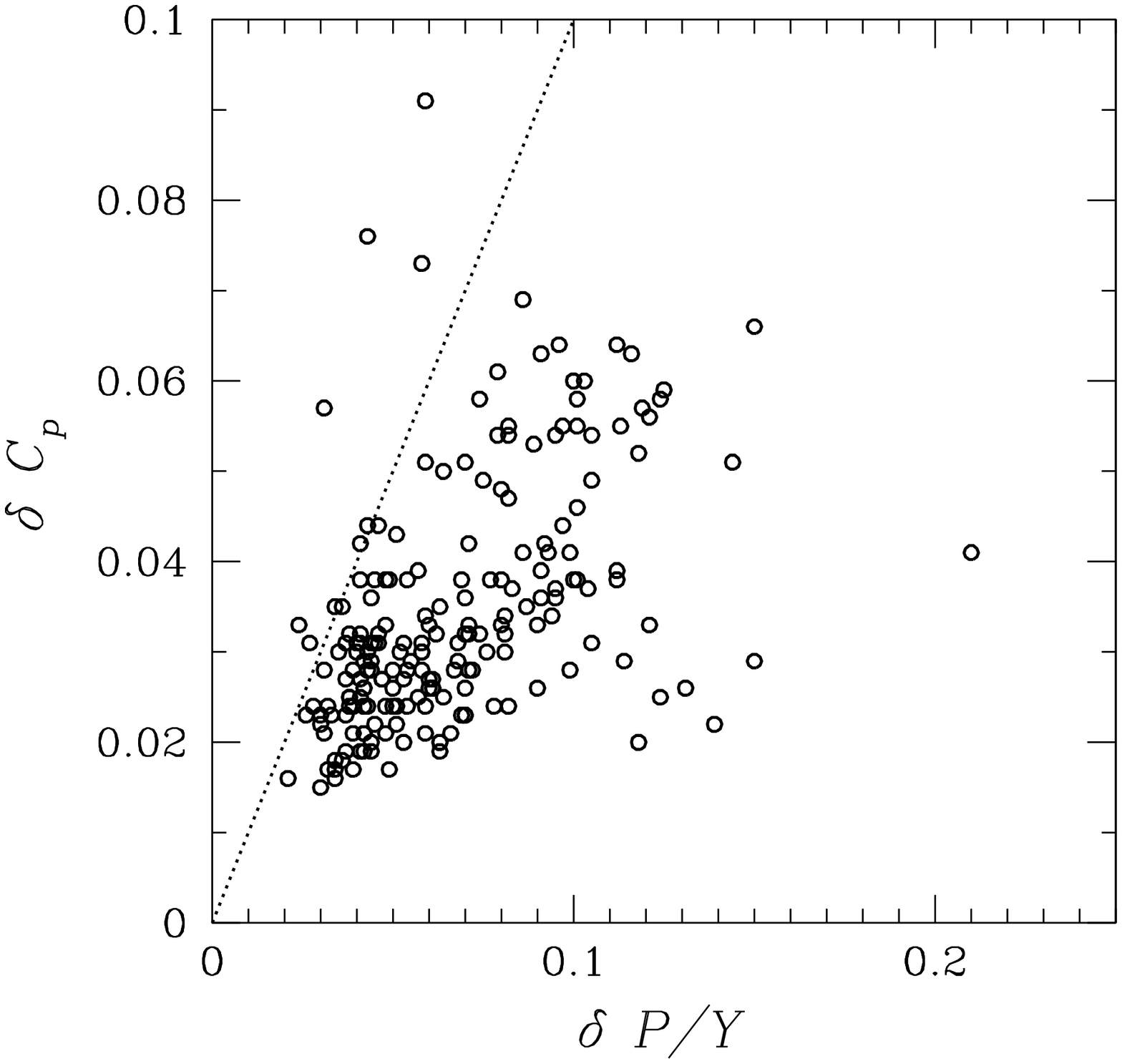}}
\caption{Correlations between the rms values of various maps with
respect to the mean trend. Each point corresponds to a single dataset from
our ensemble of 68$\times$3 simulated clusters (see
Section~\ref{s:mock_obs}). All calculations have been carried out within
0.5\,$r_{200}$. The dotted lines on all the plots show 1:1 correspondence in
rms level.}
\label{f:rms}
\end{figure*}

In addition to the similarity of the maps as traced by the correlation
estimators, we can quantify the link between the maps by examining the
degree of deviations in the map ratios. It is also advantageous to compare
the deviations seen in the maps to deviations from the average trend, as
well as their typical amplitudes. Our primary results are outlined in
Fig.~\ref{f:rms}. 

When comparing the rms level for pseudo pressure and Compton Y, one sees
that although the two scale very well, the pseudo pressure depicts the large
deviations more strongly. This explains why there is a reduction in the
correlation strength between taking the full map or only comparing the
deviations from the mean trend. A comparison between the pressure and
entropy deviations shows that although the pressure deviations are typically
larger, if all the deviations would be due to the shocks, the expected level
of entropy deviations would be limited to 0.05. The observed entropy
fluctuations are typically 10 times larger and therefore require different
explanation, such as survival of low-entropy gas during the cluster
assembly. A similar range of the fluctuations is reported for the REFLEX-DXL
cluster sample (Finoguenov et al. 2005); in Poole et al. (2007, 2008) the
amplitude of fluctuations in the maps is further linked to the merger state
of the cluster, which has also been suggested observationally by Zhang et
al. (2009).

Fig.~\ref{f:rms} reveals a good correlation between the dispersion in the
entropy and clumpiness.  The very presence of such a correlation implies that
the clumpiness of the gas is dominated by the large-scale gas displacements,
which are traced by the entropy map. Previously, we have shown that the
entropy deviations are nearly cospatial with deviations in clumpiness, as
shown by their reasonable correlation shown in Fig.~\ref{f:csr}. It implies
that to reduce the effect of clumpiness (which for example introduce a
scatter in the measurement of Hubble constant based on the X-ray and SZE
comparison), one should excise the zones most deviating in the entropy maps.

This suggestion can be further strengthened when considering the rms of the
P-to-Y ratio. First, we note that the rms in the individual pressure maps are
much higher than the rms of the P-to-Y ratio. The latter has a scatter lower
by a factor of 5, typically at the 10\% level.  To prove that clumpiness is
the reason for these residual fluctuations, we first note that a ratio
between P and Y could be written as ${T \times \sqrt{I} \over Y} = {T \times
  G \over Y \times C}$, where C is clumpiness. So we construct the
additional map, $C_p = {T \times G \over Y}$ to examine the fluctuations
associated with averaging in Compton Y maps. The scatter in P/Y is a factor
of two larger than in $C_p$, which proves that the clumpiness is the
dominant effect in producing deviations between pressure inferred from X-ray
observations and SZ signal.

\begin{figure*}
\includegraphics[width=8cm]{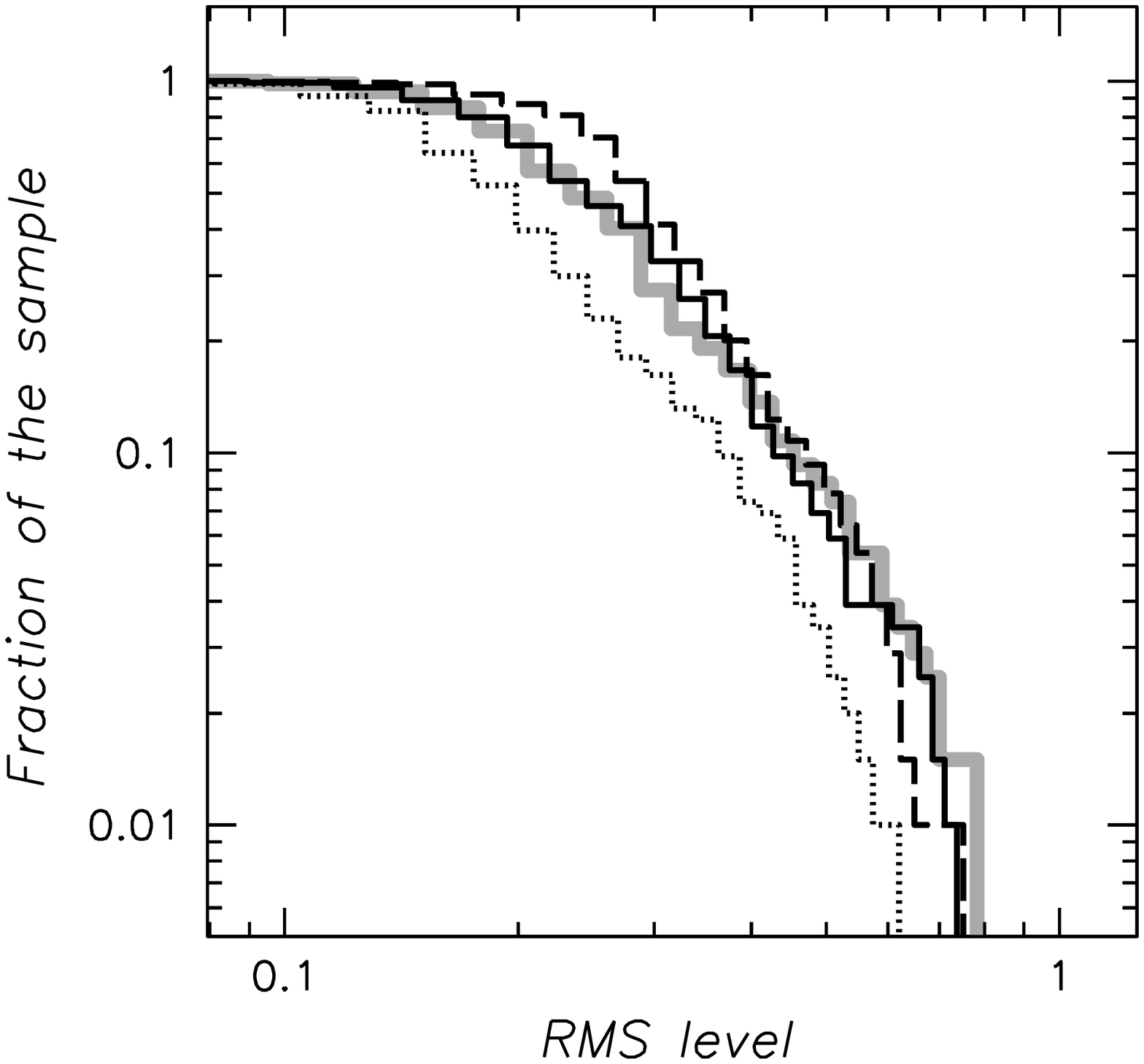}\hfill
\includegraphics[width=8cm]{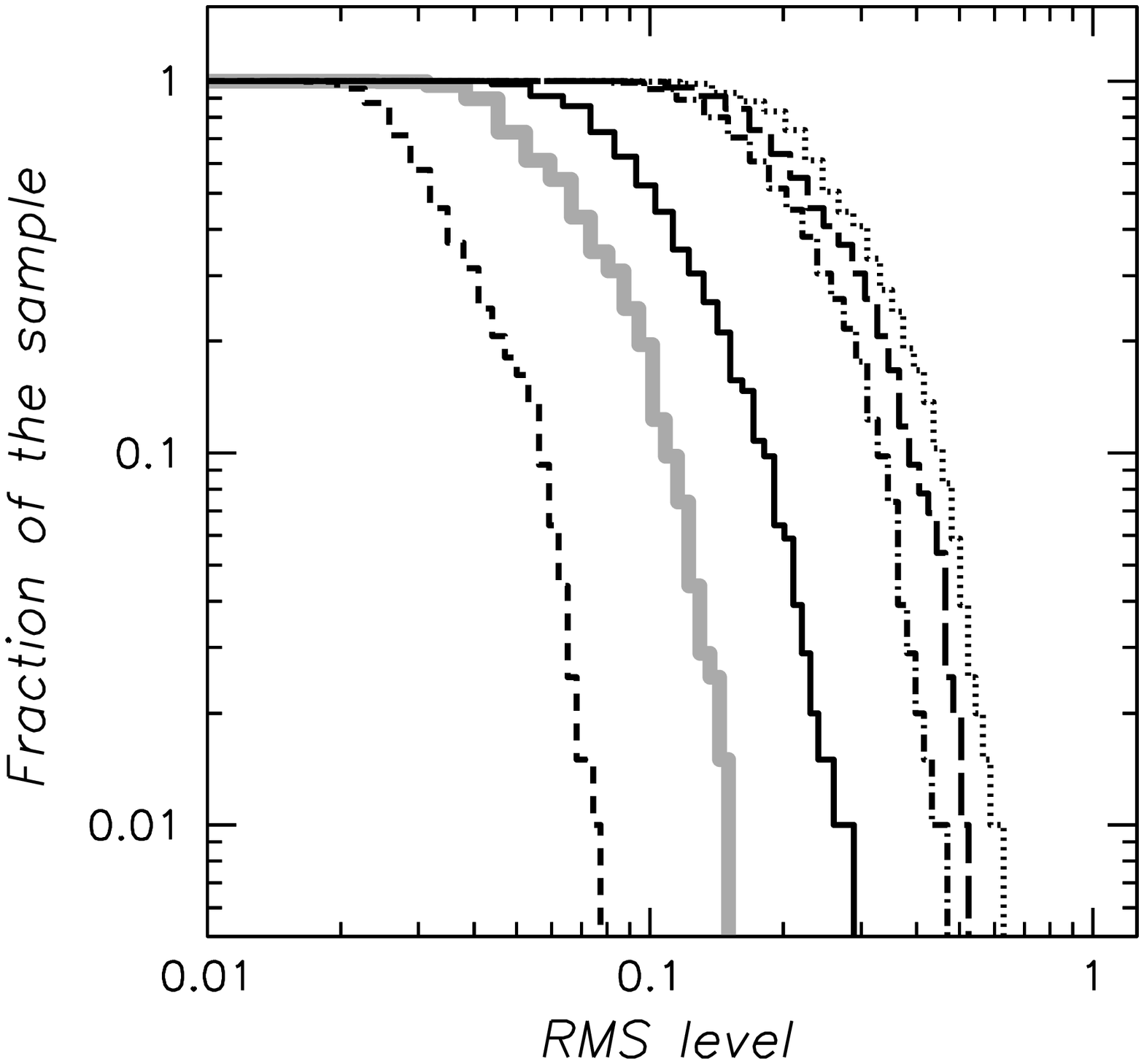}
\caption{Fraction of the sample with rms of studied parameter greater than
  the $x$-axis value. {\it Left panel:} shows pressure (grey line), square
  root of X-ray image (solid line), Compton Y (dotted line), dark matter
  (dashed line). {\it Right panel:} shows P2Y (grey), temperature (solid),
  clumpiness (dotted), $C_p$ (short-dashed), Y2I (long-dashed line),
  entropy (dot-dashed line).
\label{f:rmshst}}
\end{figure*}

To estimate the importance of certain level of deviations, we next perform
the analysis of the frequency of their occurrence.  Fig.~\ref{f:rmshst}
displays the histogram of the rms distribution for the major parameters of
this study.  Already in Finoguenov et al. (2005, 2007) we have compared the
observations to these predictions, revealing deviations in pressure
fluctuations observed on the mass scale of galaxy groups. The first
distribution of dark matter substructure has been presented in Smith et al.
(2005), manifesting that comparison with various datasets will not be
limited to X-ray observations. As evident from Fig.~\ref{f:rmshst}, the
Compton Y has the smallest deviations compared to either dark matter, X-ray
pressure or the square root of the X-ray image. The latter three exhibit
similar deviations, with the exception that for the dark matter, deviations
at the 25\% level are seen much more frequently.  As already mentioned
above, the deviations in P/Y are intermediate between the temperature and
$C_p$. Fluctuations in the clumpiness itself are much larger and the
frequency of their appearance is similar to that of entropy.

\section{Conclusions}

We have used a set of hydrodynamical  galaxy cluster simulations to
investigate the two-dimensional comparison between maps of different
quantities which are increasingly becoming available to the observer. We
test the idea of using these maps for hydrodynamical analysis and for achieving better
accuracy in using clusters for cosmological studies. We conclude that the
X-ray pseudo pressure map is a very good proxy for the Compton Y image
obtained from SZ observations. It could be used to simulate the
observational appearance of the clusters on the SZ sky, their scatter from
the averaged trend and, most importantly, provide the lowest scatter
comparison between spatially resolved SZ and X-rays measurements.

We have also shown that the pseudo entropy maps can be used as a proxy of
clumpiness of the gas, which cannot otherwise be observed directly. We
demonstrate that the scatter in the ratio of the X-ray pressure estimate to
the Compton Y is driven by clumpiness and not by shocks, so elimination of
clumpy zones, e.g. marked as deviant on the entropy maps, is the most
critical step towards achieving a consistent picture of clusters with
various methods. With the prescriptions discussed in this work, we
anticipate a reduction of scatter in the SZE-to-X-ray comparisons from 30\%
to 5\%, through the use of X-ray pressure restoration and excising strongly
deviant zones in the entropy maps.

We have shown that the structure in dark matter maps are poorly correlated
with the structure in any gas-based maps. Therefore the power of using the
gas-maps to trace the dark matter substructure is quite limited.  A degree
to which this correlation will be observed in the real data is sensitive to
the detailed physics of gas disruption and as such should allow to constrain
the role of cooling, conduction and viscosity through matching the
observations to sets of hydrodynamical simulations, as advocated in Dolag et
al. (2004).  We have also introduced a different test, which describes the
frequency of the observed rms in the maps. The has already enabled a
comparison to observations (Finoguenov et al. 2005; 2007), revealing that
while the scatter in both entropy and pressure maps in clusters agree well
with the prediction of these simulations, the amount of scatter seen in the
pressure maps of galaxy groups is much larger, indicating a larger role of
internal feedback processes there.

\begin{acknowledgements}
  AF acknowledges support from BMBF/DLR under grant 50 OR 0207 and MPG. AF
  thanks the UIUC for the hospitality during his visits. This work has been
  partially supported by a SAO grant GO8-9126B to UMBC. AEE acknowledges
  support from NASA TM4-5008X and NSF AST-0708150. The authors thank the
  anonymous referee for insightful comments on the manuscript. AF thanks
  Yu-Ying Zhang and James Taylor for their comments on the manuscript.
\end{acknowledgements}


\begin{thebibliography}{}
\bibitem{} Bialek, J.J., Evrard, A.E. \& Mohr, J.J. 2001, ApJ, 555, 597
\bibitem{} Bialek, J.J., Evrard, A.E. \& Mohr, J.J. 2002, ApJ, 587, L9
\bibitem[B{\"o}hringer et 
al.(2001)]{2001A&A...369..826B} B{\"o}hringer, H., et al.\ 2001, \aap, 369, 826
\bibitem[B{\"o}hringer et al.(2000)]{2000ApJS..129..435B} B{\"o}hringer, 
H., et al.\ 2000, \apjs, 129, 435 
\bibitem[Bonamente et al.(2006)]{2006ApJ...647...25B} Bonamente, M., Joy, 
M.~K., LaRoque, S.~J., Carlstrom, J.~E., Reese, E.~D., 
\& Dawson, K.~S.\ 2006, \apj, 647, 25 
\bibitem[Borgani \& Guzzo(2001)]{2001Natur.409...39B} Borgani, S., \& Guzzo, L.\ 2001, \nat, 409, 39 
\bibitem[Briel 
\& Henry(1994)]{1994Natur.372..439B} Briel, U.~G., \& Henry, J.~P.\ 1994, \nat, 372, 439 
\bibitem[Briel et 
al.(2004)]{2004A&A...426....1B} Briel, U.~G., Finoguenov, A., \& Henry, J.~P.\ 2004, \aap, 426, 1 
\bibitem[David et al.(1993)]{1993ApJ...412..479D} David, L.~P., Slyz, A., 
Jones, C., Forman, W., Vrtilek, S.~D., 
\& Arnaud, K.~A.\ 1993, \apj, 412, 479 
\bibitem[Edge et al.(1992)]{1992MNRAS.258..177E} Edge, A.~C., Stewart, 
G.~C., \& Fabian, A.~C.\ 1992, \mnras, 258, 177 
\bibitem[Evrard(1990)]{1990ApJ...363..349E} Evrard, A.~E.\ 1990, \apj, 363, 
349 
\bibitem[Evrard(1988)]{1988MNRAS.235..911E} Evrard, A.~E.\ 1988, \mnras, 
235, 911 
\bibitem[Finoguenov et al.(2007)]{2007MNRAS.374..737F} Finoguenov, A., 
Ponman, T.~J., Osmond, J.~P.~F., \& Zimer, M.\ 2007, \mnras, 374, 737 
\bibitem[Finoguenov et al.(2006)]{2006ApJ...646..143F} Finoguenov, A., 
Davis, D.~S., Zimer, M., \& Mulchaey, J.~S.\ 2006a, \apj, 646, 143 
\bibitem[Finoguenov et al.(2006)]{2006ApJ...643..790F} Finoguenov, A., 
Henriksen, M.~J., Miniati, F., Briel, U.~G., 
\& Jones, C.\ 2006b, \apj, 643, 790 
\bibitem[Finoguenov et 
al.(2005)]{2005A&A...442..827F} Finoguenov, A., B{\"o}hringer, H., \& Zhang, Y.-Y.\ 2005, \aap, 442, 827 
\bibitem[Finoguenov et al.(2004)]{2004ApJ...611..811F} Finoguenov, A., 
Henriksen, M.~J., Briel, U.~G., de Plaa, J., 
\& Kaastra, J.~S.\ 2004, \apj, 611, 811 
\bibitem[Dolag et al.(2004)]{2004ApJ...606L..97D} Dolag, K., Jubelgas, M., 
Springel, V., Borgani, S., \& Rasia, E.\ 2004, \apjl, 606, L97 
\bibitem[Gursky et al.(1972)]{1972ApJ...173L..99G} Gursky, H., Solinger, 
A., Kellogg, E.~M., Murray, S., Tananbaum, H., Giacconi, R., 
\& Cavaliere, A.\ 1972, \apjl, 173, L99 
\bibitem[Henry et al.(2004)]{2004ApJ...615..181H} Henry, J.~P., Finoguenov, 
A., \& Briel, U.~G.\ 2004, \apj, 615, 181 
\bibitem[Markevitch et al.(1996)]{1996ApJ...472L..17M} Markevitch, M.~L., 
Sarazin, C.~L., \& Irwin, J.~A.\ 1996, \apjl, 472, L17 
\bibitem[Markevitch(1996)]{1996ApJ...465L...1M} Markevitch, M.\ 1996, 
\apjl, 465, L1 
\bibitem[Markevitch et al.(1999)]{1999ApJ...521..526M} Markevitch, M., 
Sarazin, C.~L., \& Vikhlinin, A.\ 1999, \apj, 521, 526 
\bibitem[Mazzotta et al.(2002)]{2002ApJ...567L..37M} Mazzotta, P., Kaastra, 
J.~S., Paerels, F.~B., Ferrigno, C., Colafrancesco, S., Mewe, R., 
\& Forman, W.~R.\ 2002, \apjl, 567, L37 
\bibitem[Navarro et al.(1997)]{1997ApJ...490..493N} Navarro, J.~F., Frenk, 
C.~S., \& White, S.~D.~M.\ 1997, \apj, 490, 493 
\bibitem[Pawl et al.(2005)]{2005ApJ...631..773P} Pawl, A., Evrard, A.~E., 
\& Dupke, R.~A.\ 2005, \apj, 631, 773 
\bibitem[Powell et al.(2009)]{2009MNRAS.400..705P} Powell, L.~C., Kay, 
S.~T., \& Babul, A.\ 2009, \mnras, 400, 705 
\bibitem[Poole et al.(2007)]{2007MNRAS.380..437P} Poole, G.~B., Babul, A., 
McCarthy, I.~G., Fardal, M.~A., Bildfell, C.~J., Quinn, T., 
\& Mahdavi, A.\ 2007, \mnras, 380, 437 
\bibitem[Poole et al.(2006)]{2006MNRAS.373..881P} Poole, G.~B., Fardal, 
M.~A., Babul, A., McCarthy, I.~G., Quinn, T., 
\& Wadsley, J.\ 2006, \mnras, 373, 881 
\bibitem[Schuecker et 
al.(2004)]{2004A&A...426..387S} Schuecker, P., Finoguenov, A., Miniati, F., B{\"o}hringer, H., \& Briel, U.~G.\ 2004, \aap, 426, 387 
\bibitem[Simionescu et 
al.(2007)]{2007A&A...465..749S} Simionescu, A., B{\"o}hringer, H., Br{\"u}ggen, M., \& Finoguenov, A.\ 2007, \aap, 465, 749 
\bibitem[Smith et al.(2005)]{2005MNRAS.359..417S} Smith, G.~P., Kneib, 
J.-P., Smail, I., Mazzotta, P., Ebeling, H., 
\& Czoske, O.\ 2005, \mnras, 359, 417 
\bibitem[Sunyaev 
\& Zeldovich(1980)]{1980ARA&A..18..537S} Sunyaev, R.~A., \& Zeldovich, I.~B.\ 1980, \araa, 18, 537 
\bibitem[Vikhlinin et al.(2001)]{2001ApJ...551..160V} Vikhlinin, A., 
Markevitch, M., \& Murray, S.~S.\ 2001, \apj, 551, 160 
\bibitem[Zhang et al.(2009)]{2009ApJ...699.1178Z} Zhang, Y.-Y., Reiprich, 
T.~H., Finoguenov, A., Hudson, D.~S., 
\& Sarazin, C.~L.\ 2009, \apj, 699, 1178 
\bibitem[White(2003)]{2003ApJ...597..650W} White, M.\ 2003, \apj, 597, 650 
\end{thebibliography}
\end{document}